\documentclass[reprint,aps,prl,showpacs,twocolumn]{revtex4-1}

\usepackage[utf8]{inputenc}
\usepackage{mathptmx}
\usepackage{mathtools}
\usepackage{amsmath} 
\usepackage{braket}
\usepackage{float}
\usepackage{lipsum}  
\usepackage{svg}

\usepackage{mathptmx}
\usepackage{stackrel}
\usepackage{mathtools}
\usepackage{amsmath} 
\usepackage{braket}
\usepackage{float}
\usepackage{lipsum}  
\usepackage{svg}

\usepackage[sort&compress]{natbib}
\usepackage{mhchem}
\usepackage{changepage}
\usepackage{float}
\usepackage{mathrsfs}
\usepackage{tikz}
\usepackage{mathrsfs}
\usepackage{tikz}

\newcommand{\eref}[1]{Eq.~(\ref{#1})}%

\date{}

\begin{document}

\title{Loss of Percolation Transition in the Presence of Simple Tracer-Media Interactions}

\author{{\normalsize{}Ofek Lauber Bonomo$^{1,}$}
{\normalsize{}}}
\email{ofekzvil@mail.tau.ac.il}

\author{{\normalsize{}Shlomi Reuveni$^{1,}$}
{\normalsize{}}}
\email{shlomire@tauex.tau.ac.il}

\affiliation{\noindent \textit{$^{1}$School of Chemistry, Center for the Physics \& Chemistry of Living Systems, Ratner Institute for Single Molecule Chemistry, and the Sackler Center for Computational Molecular \& Materials Science, Tel Aviv University, 6997801, Tel Aviv, Israel}}

\date{\today}


\begin{abstract}
\noindent Random motion in disordered media is sensitive to the presence of obstacles which prevent atoms, molecules, and other particles from moving freely in space. When obstacles are static, a sharp transition between confined motion and free diffusion occurs at a critical obstacle density: the percolation threshold. To test if this conventional wisdom continues to hold in the presence of simple tracer-media interactions, we introduce the \textit{Sokoban} random walk. Akin to the protagonist of an eponymous video game, the \textit{Sokoban} has some ability to push away obstacles that block its path. While one expects this will allow the \textit{Sokoban} to venture further away, we surprisingly find that this is not always the case. Indeed, as it moves --- pushing obstacles around --- the \textit{Sokoban}  always confines itself to a finite region whose mean size is uniquely determined by the initial obstacle density. Consequently, the percolation transition is lost. This finding breaks from the ruling “ant in a labyrinth” paradigm, vividly illustrating that even weak and localized tracer-media interactions cannot be neglected when coming to understand transport phenomena.  
\end{abstract}

\maketitle

More than a century after their introduction to the readers of \emph{Nature} by Karl Pearson \cite{KarlPearson}, random walks continue to fascinate and draw attention \cite{book1,book2}. While initially motivated by the theory of gambling \cite{Gambling} and financial speculation \cite{LouisBachelier}, random walks became important in the natural sciences following the pioneering works of Einstein \cite{Einstein} Smoluchowski \cite{Smoluchowski} and others \cite{Others} on diffusion of atoms and molecules. In the time following the publication of these seminal works, random walks were further established as a versatile modelling tool \cite{RWver1,RWver2,RWver3,RWver4,RWver5,RWver6}, with applications in physics \cite{Phys1,Phys2,Phys3,Phys4,Phys5,Phys6,Phys7}, chemistry \cite{Chem1,Chem2, Chem3,Chem4,Chem5,Chem6}, biology, \cite{Biology1,Biology2,Biology3,Biology4,Biology5} movement ecology \cite{Ecology1,Ecology2,Ecology3,Ecology4,Ecology5,Ecology6}, finance and economics \cite{Economics1, Economics2}.

One paradigmatic random walk is the “ant in a labyrinth'' \cite{Ant1}, which was introduced by Pierre-Gilles de Gennes, as a simple model for diffusion in disordered media \cite{DisorderdM1,DisorderdM2,DisorderdM3}. 
Consider an ant walking on a two-dimensional square lattice, where a fraction $\rho$ of the lattice sites are occupied with obstacles, and all other sites are empty. Each time unit, the ant takes a step onto an empty neighbouring site that is chosen randomly. Given a specific lattice size and density $\rho$, one can ask how does the mean squared displacement (MSD) of the ant depend on time? When $\rho$ is small, i.e., most sites are empty, the ant's motion is almost unobstructed. In this case, the MSD scales linearly with time. On the other extreme, when $\rho$ is large, most sites are occupied by obstacles and the ant's motion is highly restricted. In this limit the ant will find itself caged in the labyrinth, resulting in an MSD that saturates asymptotically. As it turns out, for large enough systems, the transition between restricted motion and free diffusion is sharp, occurring at a critical density $0<\rho_{c}<1$ \cite{PercolationBook1}. 


\begin{figure}[t!]
\centering
\includegraphics[width=1 \linewidth]{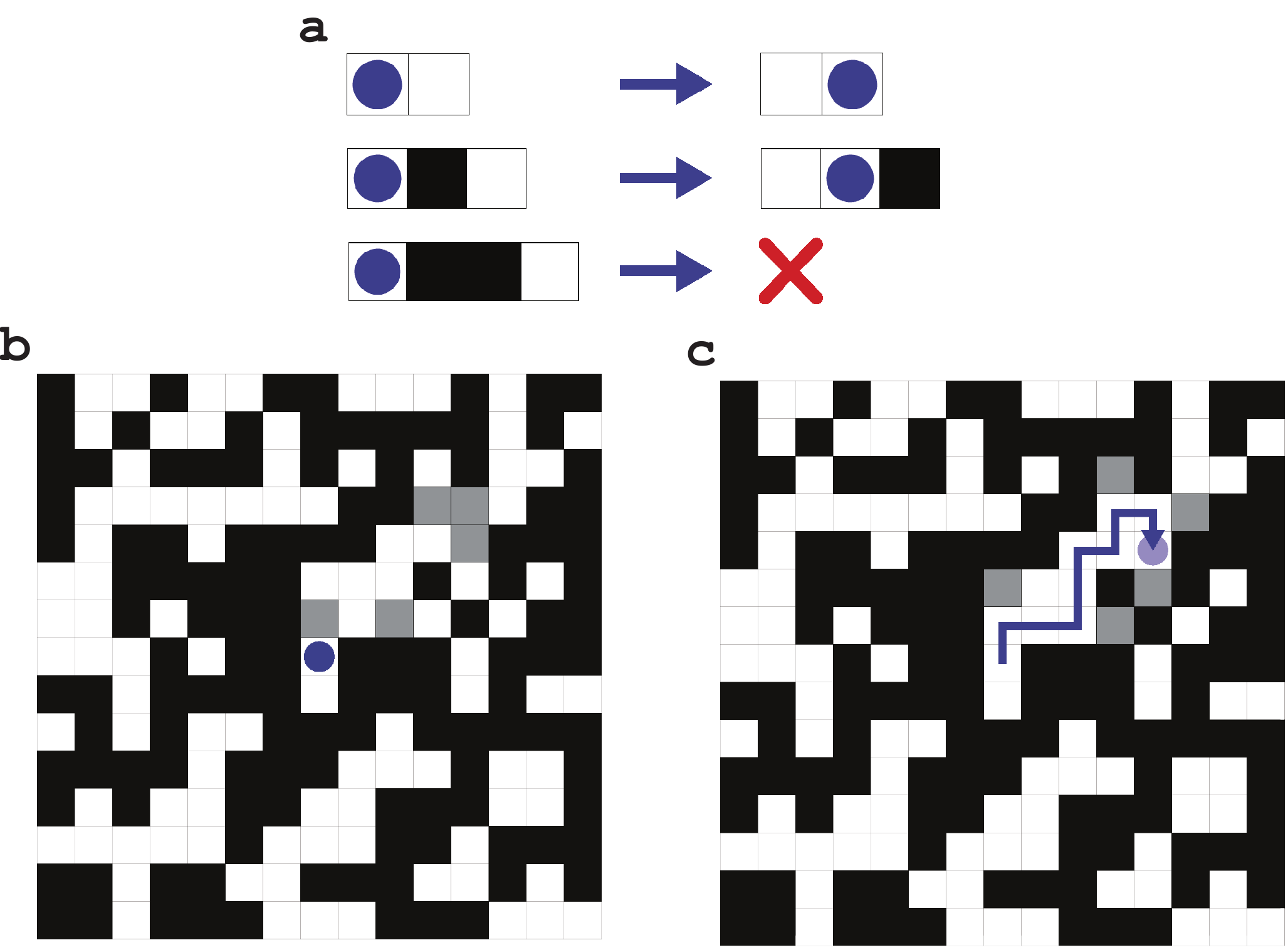}
\caption{The \textit{Sokoban} random walk. Panel (a): Laws of motion. The walk has two feasible moves: (i) it can step into an unoccupied site; (ii) it can step into an occupied site by pushing away an obstacle that occupied it one site forward, in its direction of motion. Only one obstacle can be pushed at a time. Thus, two occupied sites in a row create a block. At each time step, the walker chooses between all feasible moves with equal probability.  Panel (b): Initial configuration of a $15 \times 15$ arena. Arenas are generated by randomly distributing obstacles, such that each site has probability $\rho$ to be occupied. White squares indicate unoccupied sites. Black and gray squares indicate obstacles (these are identical, but distinguished here for clarity). Panel (c): A possible trajectory of the \textit{Sokoban} random walk. Gray squares indicate obstacles that were pushed during the course of the walk.}
\label{RulesFig}
\end{figure}


The basic assumption in de Gennes' model is that obstacles comprising the media are immobile. Namely, it is assumed that the ant's motion has no effect on the distribution of obstacles around it. Yet, this assumption is often violated, e.g. when considering active particles that burn energy and exert strong forces on their surroundings. Ironically, the ant --- which can lift many times its body weight --- is a quintessential example of such an active particle. Assuming immobile obstacles is thus fair when considering a hydrogen atom diffusing in a solid, 
but the validity of this assumption is questionable for animals and microorganisms plowing their way through crowded environments. Validity is also questionable on the microscopic scale where one may  encounter immobile obstacles that cannot be nudged by thermal fluctuations (on relevant time scales), but may nevertheless be pushed around by active particles in the media. Yet, to date, little is known on if and how transport properties are affected as a result.

\textit{The model.}---We introduce a minimalist model to show that tracer-media interactions from the type mentioned above result in a drastic, qualitative, change of transport properties. To this end, we consider a random walker that has \emph{some} ability to push away obstacles that block its path. We imagine an $n \times n$ square arena where a fraction $\rho$ of the available sites are occupied by obstacles. Taking $n$ to be odd, we place a random walker at the center of this arena. The random walk then takes place according to the following rules which are illustrated in Fig. \ref{RulesFig}a. The walker can move into an unoccupied neighbouring site, placed horizontally or vertically relative to its position. In addition, even when a site is occupied by an obstacle, the walker can move into this site while pushing the obstacle one site forward, in its direction of motion. Yet, this can only be done provided that the next site (in the direction of motion) is vacant. Thus, the walker cannot push more than one obstacle at a time. Finally, at each time step, the walker chooses between all feasible moves with equal probability.

The model presented herein is inspired by the video game \textit{Sokoban} (Japanese for warehouse keeper), which was created in 1981 by Hiroyuki Imabayashi. The premise of the game is simple: The player, playing as the keeper, pushes boxes around in a warehouse, in attempt to transport them to marked storage locations. The rules of the game are similar to the rules of the walk presented in Fig. \ref{RulesFig}. While being fairly simple to play, solving \textit{Sokoban} puzzles turns out to be a difficult computational task. It was first proved to be NP-hard \cite{NPHard} and was later shown to be PSPACE-complete \cite{PSPACE}. 

An illustration of a trajectory of the \textit{Sokoban} random walk is given in panels (b) and (c) of Fig. \ref{RulesFig}. The initial configuration of the arena is given in panel (b), and the trajectory of the walk is illustrated in panel (c). Note that a simple random walk, i.e., one that cannot push obstacles that stand in its path, would have actually been caged by the initial configuration of the arena. In contrast, the \textit{Sokoban} was able to escape this cage by pushing some of the obstacles surrounding it (highlighted in gray). More generally, we expect that the ability to push obstacles will enable the \textit{Sokoban} random walk to venture further away from its initial position when compared to a simple random walk without this pushing ability (“ant in a labyrinth'').

\begin{figure}[t!]
\includegraphics[width=0.487\textwidth]{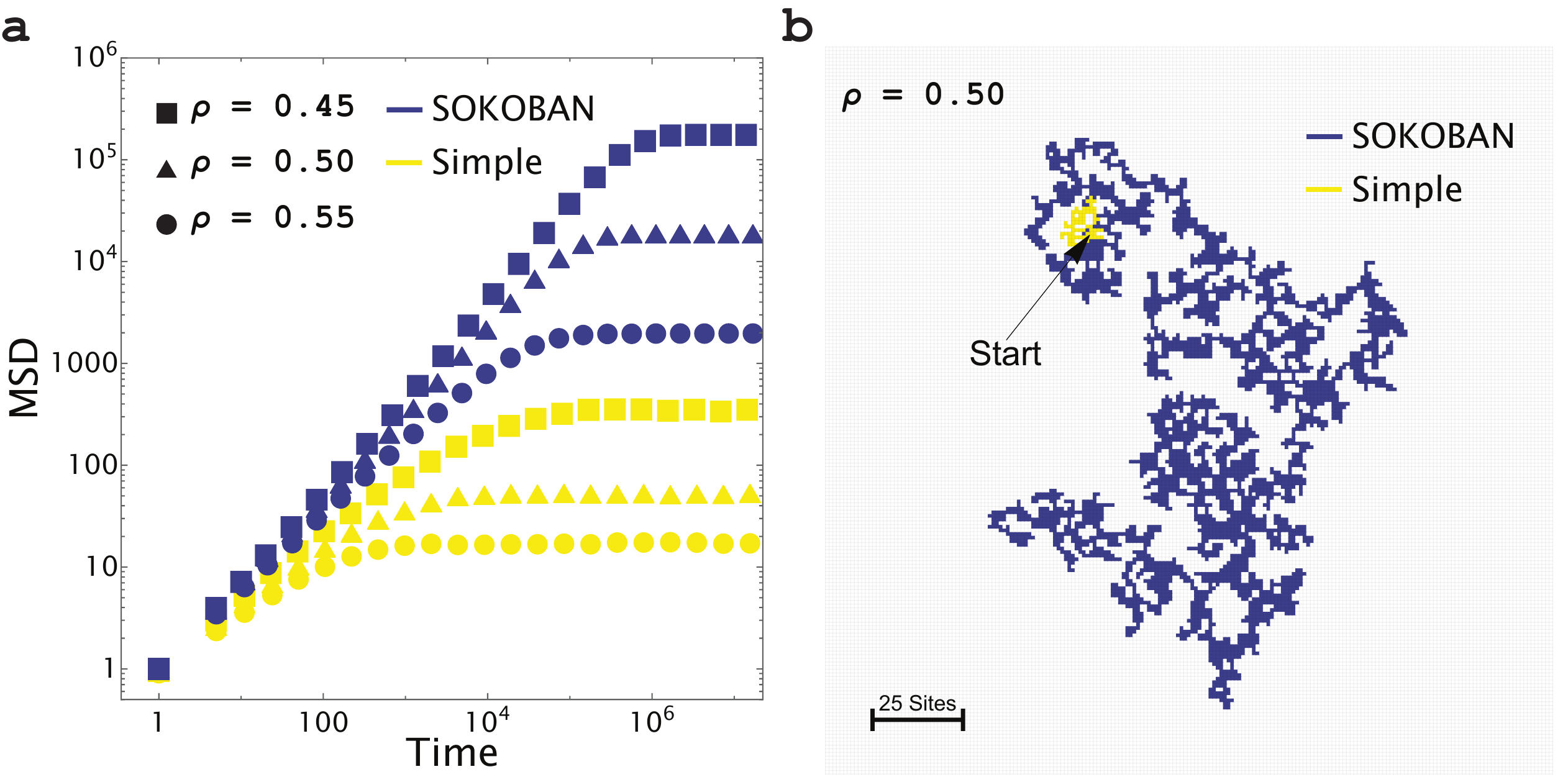}
\caption{\textit{Sokoban} vs. simple random walk. Panel (a): Mean squared displacement (MSD) of a simple random walk (yellow) and the \textit{Sokoban} random walk (purple) as a function of time, for three different obstacle densities: $\rho = 0.45, 0.5, 0.55$. In the long time limit, the MSD of the \textit{Sokoban} is orders of magnitude higher compared to the MSD of the simple random walk. Panel (b): Trajectories of the simple (yellow) and \textit{Sokoban} (purple) random walks, starting in an \emph{identical} arena with $\rho = 0.5$. The difference in MSD is evident.}
\label{MSDFig}
\end{figure}

\textit{Monte Carlo simulations.}---To test this hypothesis, we simulate the \textit{Sokoban} and simple random walks, for a large number of randomly generated and sufficiently large arenas, so as to completely avoid boundary effects. In Fig. \ref{MSDFig}a we plot the mean squared displacement, given by $MSD(t) = \langle r^2(t) \rangle$, where $r(t)$ is the Euclidean distance to the initial position at time $t$, and $\langle \cdot \rangle$ indicates an ensemble average over all generated walks. Plots are made for the \textit{Sokoban} (purple) and simple (yellow) random walks at three different obstacle  densities. As expected, in the long time limit, the MSD of the \textit{Sokoban} is significantly higher compared to the MSD of the simple random walk. As a result, the \textit{Sokoban}  explores larger portions of the arena as illustrated by the trajectories given in Fig. \ref{MSDFig}b. Further illustration of the \textit{Sokoban} walk is provided in a supplementary video (SV1.avi).


\begin{figure*}[t!]
\includegraphics[width=1\textwidth]{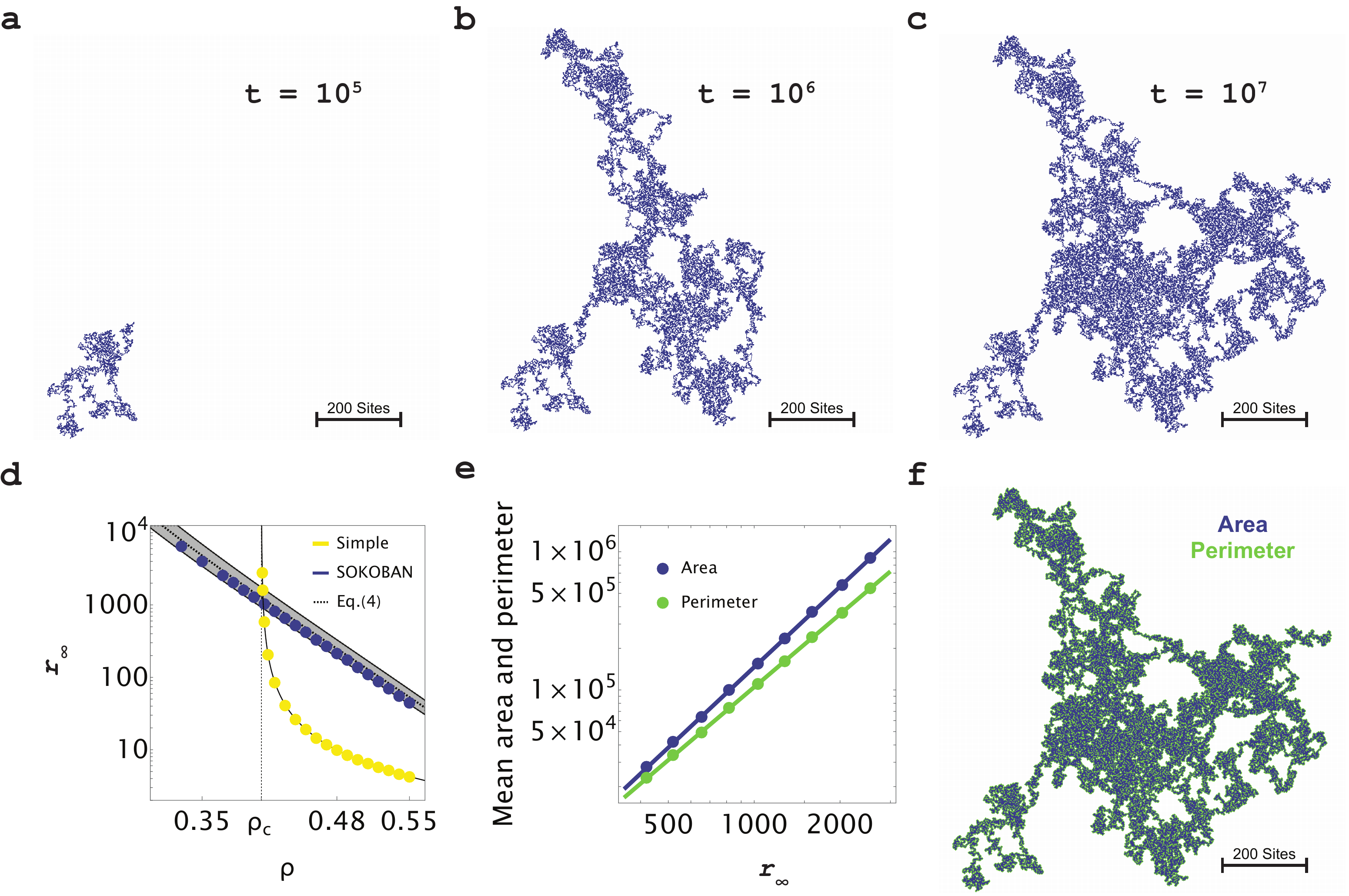}
\centering
\caption{Loss of percolation transition in the \textit{Sokoban} random walk. Panels (a-c): Sites visited by the \textit{Sokoban} random walk, in an arena with an obstacle density of $\rho = 0.4$, for $t= 10^5, 10^6, 10^7$. By time $t=10^7$ the \textit{Sokoban} random walk is effectively caged and does not visit new sites (see comparison with $t=10^8$ in Fig. S1 \cite{SI}). Panel (d): The exploration radius  $r_{\infty}=\lim_{t\to\infty} \sqrt{MSD(t)}$ for the \textit{Sokoban} (purple markers) and simple (yellow markers) random walks vs. the obstacle density $\rho$ (log-log scale). Markers coming from simulations show that while the exploration radius of the simple random walk diverges as $\rho\to\rho _c$, this radius remains finite for the \textit{Sokoban} random walk. Simulations are in excellent agreement with \eref{scaling} whose prediction is given by a dashed line surrounded by a grayed out sleeve which indicates an error of $\pm 2.5\%$ in the estimates made for the parameters $\gamma$ and $C$. Panel (e): The mean number of sites $\langle \mathcal{A} \rangle$ visited by the  \textit{Sokoban} and mean number of perimeter sites $\langle \mathcal{P} \rangle$  follow the power-law relations of Eqs. (\ref{Mean_A}) and (\ref{Mean_P}) [note the log-log scale, also see panel (f)]. Markers come from simulations and best fits yield the following estimates of the relevant parameters $A\simeq0.2277,B\simeq0.5736,\alpha\simeq1.936,$ and $\beta\simeq1.754$. Panel (f): Area (purple) and perimeter (green) of the trajectory from  panel (c). The area of the trajectory is defined as the number of visited sites. The perimeter of the trajectory is taken as the double layer of unvisited sites that surround area sites (share a mutual edge).}
\label{TrajectoriesFig}
\end{figure*}

All densities in Fig. \ref{MSDFig} were taken to be above the 2D site-percolation threshold, namely, the critical density $\rho_{c} \simeq 0.407$ \cite{PercolationCrit}, above which the simple random walk eventually becomes restricted (caged). The fact that for these densities the \textit{Sokoban} was able to explore larger portions of the arena, hints that the critical density for this walk should be higher than, or equal to, the percolation threshold; allowing the \textit{Sokoban} to roam unbounded when the density of obstacles drops below $\rho_c$. However, when simulating the \textit{Sokoban} for $\rho < \rho_c$, we surprisingly find that its MSD still saturates in the long time limit. An example is given in panels (a-c) of Fig. \ref{TrajectoriesFig}, where we take $\rho =0.4$, and present a time evolution of a typical \textit{Sokoban} trajectory. Snapshots are taken for $t = 10^5, 10^6$ and $10^7$. For $t \gtrsim 10^7$ the walk does not visit new sites, indicating it is indeed confined (see Fig. S1 \cite{SI}). 

Further evidence that the \textit{Sokoban} random walk dynamically confines itself at densities lower than the percolation threshold comes from extensive numerical simulations that we perform for this system. Defining the exploration radius  $r_{\infty}=\lim_{t\to\infty} \sqrt{MSD(t)}$, i.e., the level at which the square root of the MSD saturates, we plot this quantity as a function of $\rho$ for the \textit{Sokoban} and simple random walks (Fig. \ref{TrajectoriesFig}d). As expected, for the simple random walk $r_{\infty}$ diverges when $\rho$ approaches $\rho _c \simeq 0.407$ from above; indicating the existence of a critical density  beyond which the walk is no longer confined. However, for the \textit{Sokoban} random walk we find that $r_{\infty}$ is finite for all values of $\rho$ sampled. 

The results presented in panels (a-d) of Fig. \ref{TrajectoriesFig} assert that the critical density of the \textit{Sokoban} random walk cannot be higher than the percolation threshold. Thus, if such critical density even exists, it must be lower than the percolation threshold. Alternatively, it is possible that the \textit{Sokoban} random walk does not have a critical density and that this walk dynamically confines itself at every positive obstacle density $\rho>0$. One way to try and find out will be to simulate this system for increasingly smaller obstacle densities, which in turn requires increasingly larger arenas. However, this brute force approach is limited computationally and quickly runs into trouble.

\textit{Scaling approach.}---Instead, we take a scaling approach seeking better physical understanding for why the \textit{Sokoban} random walk might be dynamically confining itself. We first note that in order for confinement to occur the \textit{Sokoban} must push surrounding obstacles until it eventually creates a cage from which it cannot escape, e.g. see Fig. \ref{RulesFig}c, and  supplementary video (SV1.avi). Visited sites within the cage will be surrounded by a double-layer of obstacles (perimeter) that prevents the walker from accessing additional sites. This double-layer is required since  the \textit{Sokoban} will otherwise be able to push its way out and breach the perimeter. 

To further proceed, we formulate a simple condition that leads to caging. We define $\mathcal{A}$ to be the area covered by the \textit{Sokoban} trajectory, i.e., the total number of sites visited in the long time limit. In addition, we let $\mathcal{P}$ stand for the number of sites in the double-layered perimeter that surrounds visited sites. For example, the double-layered perimeter of the sites visited by the trajectory in Fig. \ref{TrajectoriesFig}c is indicated in Fig. \ref{TrajectoriesFig}f. Recall that area and perimeter sites were initially occupied by obstacles with the same probability $\rho$. Thus, on average, $\mathcal{A} \rho$ area sites were initially occupied while $\mathcal{P} (1-\rho)$ perimeter sites were vacant. Now, to guarantee caging, we demand
\begin{equation}
\mathcal{A} \rho  = \mathcal{P} (1-\rho). \label{mass_balance}
\end{equation}
In other words, \eref{mass_balance} states that the \textit{Sokoban} will surely get caged when  $\mathcal{A} \rho$  obstacles are pushed clear from its path to occupy the  $\mathcal{P} (1-\rho)$ perimeter sites that were initially empty. 

Due to the inherent randomness of the \textit{Sokoban} walk the $\mathcal{A}$ and $\mathcal{P}$ defined above are random, taking different values with every realization. However, averaging over many realizations we observe that these quantities obey a power-law scaling 
\begin{equation}
\langle \mathcal{A} \rangle \simeq Ar_{\infty}^{\alpha}, \label{Mean_A} 
\end{equation}
and   
\begin{equation}
\langle \mathcal{P} \rangle \simeq Br_{\infty}^{\beta}, \label{Mean_P} 
\end{equation} 
as shown in Fig. \ref{TrajectoriesFig}e. Underlying these relations is what seems like a fractal shape of the \textit{Sokoban's}  trajectories (Fig. S2 \cite{SI}). Substituting $\mathcal{A}$ and $\mathcal{P}$ in \eref{mass_balance} by their averages and rearranging we obtain 
\begin{equation} 
r_{\infty} = \left ( \frac{1-\rho}{C\rho} \right )^{1/\gamma}, \label{scaling} 
\end{equation}

\noindent where $\gamma = \alpha-\beta$ and $C = A/B$.

Equation (\ref{scaling}) conveys a relation between the exploration radius, $r_{\infty}$, of the \textit{Sokoban} random walk and the obstacle density $\rho$. To test it, we fit the simulations data in Fig. \ref{TrajectoriesFig}e and estimate the parameters $\{A,B,\alpha,\beta\}$ which govern the power-law scalings of $\langle \mathcal{A} \rangle$ in \eref{Mean_A} and $\langle \mathcal{P} \rangle$ in \eref{Mean_P}. Using these estimates we obtain  $\gamma \simeq 0.182$ and $C \simeq0.397$. Substituting these numbers back into \eref{scaling}, we plot the predicted relation (dashed line, Fig. \ref{TrajectoriesFig}d). We compare this prediction to direct estimates of the mean exploration radii that were obtained from the asymptotic MSDs at different obstacle densities (markers, Fig. \ref{TrajectoriesFig}d). Very good agreement is found between the prediction of \eref{scaling} and data coming from simulations.

While the agreement between \eref{scaling} and data coming from simulations is very good, it is not perfect. One source of error comes from the conservative assumption that was made while writing \eref{mass_balance}. Namely, that the \textit{Sokoban} gets caged only when all visited (area) sites are empty and all perimeter sites are occupied. Yet, we find that caging usually occurs earlier, with some visited sites still occupied by obstacles and some perimeter sites still empty. This can happen as the \textit{Sokoban} may get trapped in a small micro-environment that becomes isolated from the rest of the arena after caging occurs. However, modifying \eref{mass_balance} to state that the \textit{Sokoban} gets caged when a fraction $f_\mathcal{A}$ of the obstacles that resided in visited sites were pushed to occupy a fraction $f_\mathcal{P}$ of perimeter sites that were initially vacant, $\mathcal{A} f_\mathcal{A} \rho  = \mathcal{P} f_\mathcal{P} (1-\rho)$, yields $C=Af_\mathcal{A}/Bf_\mathcal{P}$ in \eref{scaling} and does not change $\gamma$. 

Interestingly, for the obstacle densities examined here, we find that the average fraction $f_\mathcal{A}$ is only slightly larger than $f_\mathcal{P}$, thus explaining the slight overestimate in the theoretical prediction of $r_{\infty}$ compared with simulations data (Fig. \ref{TrajectoriesFig}d). Whether larger deviations from \eref{scaling} arise for smaller obstacle densities is currently unknown; but cannot be entirely ruled out since $f_\mathcal{A}$ and $f_\mathcal{P}$ also show some dependence on $\rho$. 

\textit{Discussion and outlook.}---In this paper we introduced a new model for random walks in disordered media. Contrary to the canonical “ant in a labyrinth'' model, the \textit{Sokoban} random walk considered herein actively interacts and modifies its surroundings by pushing obstacles in its course of motion. We studied the dynamics of the \textit{Sokoban} using extensive Monte-Carlo simulations, measured its MSD, and compared it to that obtained for a simple random walk in the presence of obstacles. At obstacle densities above the percolation threshold, we find that the \textit{Sokoban} typically roams much further than a simple random walk that cannot push away obstacles that block its path. However, at obstacle densities that are close to the percolation threshold and lower, there is a striking change of trend: while the simple random walk becomes unbounded, the \textit{Sokoban} random walk remains confined (caged). 

A conservative rule regarding the onset of caging was used, in tandem with fractal scaling laws, to derive \eref{scaling} which relates the density of obstacles to the asymptotic root MSD of the \textit{Sokoban}. This equation explained the observed density dependence of the mean exploration radius. A prime corollary coming from \eref{scaling} is that the exploration radius remains finite for any positive obstacle density $\rho>0$, suggesting that the \textit{Sokoban} undergoes dynamical caging at all obstacle densities. Consequently, the percolation transition is lost. However, \eref{scaling} is not exact, and numerical determination of the mean exploration radius at extremely low obstacle densities is very challenging computationally and beyond our reach. Thus, the existence of a critical density in the \textit{Sokoban} model, or lack of it thereof, remains to be proven rigorously.

Despite their superficial similarity, the \textit{Sokoban} and simple random walks exhibit qualitatively different behaviours. While the simple random walk becomes unbounded below a critical obstacle density, the \textit{Sokoban} random walk undergoes dynamical caging well beyond this density. From this we learn that the ability to push away obstacles is not always beneficial for a random walker seeking to explore its surroundings. Indeed, depending on the obstacle density,  the asymptotic MSDs of the \textit{Sokoban} and simple random walks may differ by orders of magnitude. However, for a very narrow range of densities near the percolation threshold the asymptotic MSDs are similar, thus making it difficult to discriminate the two walks based solely on this static measure. To this end, we recall that near criticality a simple random walk on a percolation cluster displays sub-diffusive behaviour \cite{MSDPercolation}. In contrast, we find that the  \textit{Sokoban} displays regular diffusion, i.e., MSDs that grow linearly with time (Fig. S3 \cite{SI}). 

Short-ranged tracer-media interactions are often neglected as they are not believed to significantly impact transport properties at the macro scale. However, the findings presented herein vividly demonstrate that even a limited  ability of a random walker to dynamically modify its local environment, could drastically alter its long-ranged transport behaviour. In such cases, where strong deviations from the inert “ant in a labyrinth'' paradigm occur, the \textit{Sokoban} provides an alternative null model. Depending on the system at hand, this model can be further adapted and refined to capture essential details that may have been absent from the treatment and discussion presented herein. \\


{}


\begin{thebibliography}{99}


\bibitem{KarlPearson} Pearson, K., 1905. The problem of the random walk. Nature, 72(1865), pp.294-294.


\bibitem{book1} Lawler, G.F. and Limic, V., 2010. Random walk: a modern introduction (Vol. 123). Cambridge University Press.

\bibitem{book2} Klafter, J. and Sokolov, I.M., 2011. First steps in random walks: from tools to applications. OUP Oxford.

\bibitem{Gambling} David, F.N., 1998. Games, gods, and gambling: A history of probability and statistical ideas. Courier Corporation.

\bibitem{LouisBachelier} Bachelier, L., 1900. Theory of speculation. In Scientific Annals of the École Normale Supérieure (Vol. 17, pp. 21-86).

\bibitem{Einstein} Einstein, A., 1905. Über die von der molekularkinetischen Theorie der Wärme geforderte Bewegung von in ruhenden Flüssigkeiten suspendierten Teilchen. Annalen der physik, 4.

\bibitem{Smoluchowski} Von Smoluchowski, M., 1906. Zur kinetischen theorie der brownschen molekularbewegung und der suspensionen. Annalen der physik, 326(14), pp.756-780.

\bibitem{Others} Ebeling, W., Gudowska-Nowak, E. and Sokolov, I.M., 2008. On Stochastic Dynamics in Physics --- Remarks on History and Terminology. Acta Physica Polonica B, 39(5).



\bibitem{RWver1} Montroll, E.W. and Weiss, G.H., 1965. Random walks on lattices. II. Journal of Mathematical Physics, 6(2), pp.167-181.

\bibitem{RWver2} Kenkre, V.M., Montroll, E.W. and Shlesinger, M.F., 1973. Generalized master equations for continuous-time random walks. Journal of Statistical Physics, 9(1), pp.45-50.



\bibitem{RWver3}Schütz, G.M. and Trimper, S., 2004. Elephants can always remember: Exact long-range memory effects in a non-Markovian random walk. Physical Review E, 70(4), p.045101.


\bibitem{RWver4} Gabel, A. and Redner, S., 2012. Random walk picture of basketball scoring. Journal of Quantitative Analysis in Sports, 8(1).

\bibitem{RWver5} Giuggioli, L., 2020. Exact spatiotemporal dynamics of confined lattice random walks in arbitrary dimensions: a century after smoluchowski and pólya. Physical Review X, 10(2), p.021045.

\bibitem{RWver6} Barkai, E. and Burov, S., 2020. Packets of diffusing particles exhibit universal exponential tails. Physical review letters, 124(6), p.060603.

\bibitem{Phys1}
Scher, H. and Montroll, E.W., 1975. Anomalous transit-time dispersion in amorphous solids. Physical Review B, 12(6), p.2455.

\bibitem{Phys2} Metzler, R. and Klafter, J., 2000. The random walk's guide to anomalous diffusion: a fractional dynamics approach. Physics reports, 339(1), pp.1-77.

\bibitem{Phys3} Berkowitz, B., Cortis, A., Dentz, M. and Scher, H., 2006. Modeling non‐Fickian transport in geological formations as a continuous time random walk. Reviews of Geophysics, 44(2).

\bibitem{Phys4} Condamin, S., Bénichou, O., Tejedor, V., Voituriez, R. and Klafter, J., 2007. First-passage times in complex scale-invariant media. Nature, 450(7166), pp.77-80.

\bibitem{Phys5} Krapivsky, P.L., Redner, S. and Ben-Naim, E., 2010. A kinetic view of statistical physics. Cambridge University Press.

\bibitem{Phys6} Bray, A.J., Majumdar, S.N. and Schehr, G., 2013. Persistence and first-passage properties in nonequilibrium systems. Advances in Physics, 62(3), pp.225-361.

\bibitem{Phys7}Metzler, R., Jeon, J.H., Cherstvy, A.G. and Barkai, E., 2014. Anomalous diffusion models and their properties: non-stationarity, non-ergodicity, and ageing at the centenary of single particle tracking. Physical Chemistry Chemical Physics, 16(44), pp.24128-24164.

\bibitem{Phys8} Zaburdaev, V., Denisov, S. and Klafter, J., 2015. Lévy walks. Reviews of Modern Physics, 87(2), p.483.

\bibitem{Chem1}
Szabo, A., Lamm, G. and Weiss, G.H., 1984. Localized partial traps in diffusion processes and random walks. Journal of statistical physics, 34(1), pp.225-238.

\bibitem{Chem2}
Kang, K. and Redner, S., 1985. Fluctuation-dominated kinetics in diffusion-controlled reactions. Physical Review A, 32(1), p.435.


\bibitem{Chem3}
Froemberg, D. and Sokolov, I.M., 2008. Stationary fronts in an A+ B→ 0 reaction under subdiffusion. Physical review letters, 100(10), p.108304.

\bibitem{Chem4}
Bénichou, O., Chevalier, C., Klafter, J., Meyer, B. and Voituriez, R., 2010. Geometry-controlled kinetics. Nature chemistry, 2(6), pp.472-477.

\bibitem{Chem5}
Lanoiselée, Y., Moutal, N. and Grebenkov, D.S., 2018. Diffusion-limited reactions in dynamic heterogeneous media. Nature communications, 9(1), pp.1-16.

\bibitem{Chem6}
Scher, Y. and Reuveni, S., 2021. Unified Approach to Gated Reactions on Networks. Physical Review Letters, 127(1), p.018301.


\bibitem{Biology1} Berg, H.C., 2018. Random walks in biology. In Random Walks in Biology. Princeton University Press.


\bibitem{Biology2} Amir, A., 2014. Cell size regulation in bacteria. Physical review letters, 112(20), p.208102.

\bibitem{Biology3} Iyer‐Biswas, S. and Zilman, A., 2016. First‐passage processes in cellular biology. Advances in chemical physics, 160, pp.261-306.


\bibitem{Biology4}
Bénichou, O., Bhat, U., Krapivsky, P.L. and Redner, S., 2018. Optimally frugal foraging. Physical Review E, 97(2), p.022110.

\bibitem{Biology5}
Meyer, H. and Rieger, H., 2021. Optimal non-Markovian search strategies with n-step memory. Physical Review Letters, 127(7), p.070601.

\bibitem{Ecology1} Boyer, D., Ramos-Fernández, G., Miramontes, O., Mateos, J.L., Cocho, G., Larralde, H., Ramos, H. and Rojas, F., 2006. Scale-free foraging by primates emerges from their interaction with a complex environment. Proceedings of the Royal Society B: Biological Sciences, 273(1595), pp.1743-1750.

\bibitem{Ecology2}Sims, D.W., Southall, E.J., Humphries, N.E., Hays, G.C., Bradshaw, C.J., Pitchford, J.W., James, A., Ahmed, M.Z., Brierley, A.S., Hindell, M.A. and Morritt, D., 2008. Scaling laws of marine predator search behaviour. Nature, 451(7182), pp.1098-1102.

\bibitem{Ecology3}
Giuggioli, L., Potts, J.R. and Harris, S., 2011. Animal interactions and the emergence of territoriality. PLoS computational biology, 7(3), p.e1002008.

\bibitem{Ecology4}
Viswanathan, G.M., Da Luz, M.G., Raposo, E.P. and Stanley, H.E., 2011. The physics of foraging: an introduction to random searches and biological encounters. Cambridge University Press.


\bibitem{Ecology5}Song, C., Koren, T., Wang, P. and Barabási, A.L., 2010. Modelling the scaling properties of human mobility. Nature physics, 6(10), pp.818-823.


\bibitem{Ecology6}
Vilk, O., Orchan, Y., Charter, M., Ganot, N., Toledo, S., Nathan, R. and Assaf, M., 2022. Ergodicity breaking in area-restricted search of avian predators. Physical Review X, 12(3), p.031005.

\bibitem{Economics1}
Bouchaud, J.P. and Potters, M., 2003. Theory of financial risk and derivative pricing: from statistical physics to risk management. Cambridge university press.

\bibitem{Economics2}
Bouchaud, J.P., Bonart, J., Donier, J. and Gould, M., 2018. Trades, quotes and prices: financial markets under the microscope. Cambridge University Press.


\bibitem{Ant1} de Gennes, P.G., 1976. La percolation: un concept unificateur. La recherche, 7(72), pp.919-927.



\bibitem{DisorderdM1}Havlin, S. and Ben-Avraham, D., 1987. Diffusion in disordered media. Advances in physics, 36(6), pp.695-798.

\bibitem{DisorderdM2}Bouchaud, J.P. and Georges, A., 1990. Anomalous diffusion in disordered media: statistical mechanisms, models and physical applications. Physics reports, 195(4-5), pp.127-293.

\bibitem{DisorderdM3}
Sokolov, I.M., 2012. Models of anomalous diffusion in crowded environments. Soft Matter, 8(35), pp.9043-9052.




\bibitem{PercolationBook1} Ben-Avraham, D. and Havlin, S., 2000. Diffusion and reactions in fractals and disordered systems. Cambridge university press.



\bibitem{NPHard} Dor, D. and Zwick, U., 1999. SOKOBAN and other motion planning problems. Computational Geometry, 13(4), pp.215-228.

\bibitem{PSPACE} Culberson, J., 1997. Sokoban is PSPACE-complete.



\bibitem{PercolationCrit} Jacobsen, J.L., 2015. Critical points of Potts and O (N) models from eigenvalue identities in periodic Temperley–Lieb algebras. Journal of Physics A: Mathematical and Theoretical, 48(45), p.454003.

\bibitem{SI} See Supplemental Material for supplementary figures.

\bibitem{MSDPercolation} Gefen, Y., Aharony, A. and Alexander, S., 1983. Anomalous diffusion on percolating clusters. Physical Review Letters, 50(1), p.77.



\end{thebibliography}
\end{document}